\documentclass[onecolumn,
nobibnotes,
amsmath,amssymb,
pra,
]{revtex4-2}
\newcommand{\quotes}[1]{``#1''}

\usepackage{textcomp}
\usepackage[T1]{fontenc}
\usepackage{natbib}
\usepackage[left= 2cm,right = 2cm, bottom = 2 cm]{geometry}
\usepackage{mathtools}
\usepackage{csvsimple,array}
\addcontentsline{toc}{section}{Acknowledgement}
\usepackage{xurl}
\usepackage[utf8]{inputenc}
\usepackage{newunicodechar}
\usepackage[T1]{fontenc}
\usepackage{textcomp}
\usepackage{bbold}
\usepackage{subcaption}
\usepackage{tabularx}
\usepackage{lipsum}
\usepackage{lineno}
\modulolinenumbers[10]
\usepackage{hyperref}
\usepackage{braket}
\usepackage{dsfont}
\usepackage{braket}
\usepackage{graphicx}
\usepackage{dcolumn}

\usepackage{amsmath}
\usepackage{bm}
\usepackage[ruled,vlined]{algorithm2e}
\usepackage{euler}

\usepackage{amsmath,amsfonts}

\begin{document}

\preprint{APS/123-QED}

\title{Preparing Maximally Entangled States By Monitoring the Environment-System Interaction In Open Quantum Systems }

\author{Ali A. Abu-Nada}
 
\affiliation{
 Faculty of Engineering Technology and Science, Higher Colleges of Technology, Abu Dhabi, UAE }
 
 \author{Moataz A. Salhab}
 \affiliation{
Faculty of Information  Technology, Higher Colleges of Technology, Abu Dhabi, UAE }

\date{\today}

\begin{abstract}

 A common assumption in open quantum systems in general is that the noise induced by the environment, due to  the continuous  interaction between a quantum system and its environment, is responsible for the  disappearance of quantum properties of this  quantum system. Interestingly, we show that an environment can be engineered  and controlled to direct  an arbitrary  quantum  system towards a maximally entangled state and thus  can be considered  as a resource for quantum information processing. Barreiro $et. al.$ [Nature 470, 486 (2011)] demonstrated this idea  experimentally using an open-system quantum simulator up to five  trapped ions .\\
 In this paper, we direct an arbitrary initial mixed state of two and four qubits,   which is interacting with its  environment, into a  maximally entangled state . We use  QASM simulator and also an IBM Q real processor, with and  without errors mitigating,  to investigate the effect of the noise on  the preparation of the initial mixed state of the  qubits in addition to the population of the target  state.  

\end{abstract}

\maketitle

\section{Introduction}
An elementary requirement in the open quantum system is the perfect isolation of the system from its environment to avoid decoherence. The decoherence causes, for example, spontaneous emission on the two level system quantum system (qubit), and this is describing as the fundamental quantum noise. On the other hand, any quantum computer requires a non-reversible and therefore  incoherent operations for state initialization and measurements \cite{nielsen}. For instance, quantum error correction techniques depend on controlled incoherent operations through   an algorithm to remove errors from the system similar to state initialization \cite{phdthesis-full}. Interestingly, it has been shown theoretically \cite{verstraete,zhou} and experimentally \cite{barreiro,Schindler} that non-coherent operations can be  designated as a resource for quantum information. Indeed, these ideas can only be implemented if controlled non-coherent operations are available in the system. Mathematically, these non-reversible operations are defined  by a trace-preserving completely positive map $\mathcal{E}(\rho)$ \cite{nielsen} acting on a density matrix instead of  unitary operations acting on pure states. On the other  hand, one of the  crucial elementary part of the  quantum information processing is preparing a certain quantum state and retaining it for an appropriately relatively long time \cite{PhysRevLett.89.270403}. An encouraging viewpoint in generating quantum states with favorite properties is obtainable by using a controlled dissipation. With the assistance of the controlled dissipation, one can prepare and preserve maximally entangled qubit states from arbitrary initial mixed state of $n$-qubits and correlated quantum phases from this arbitrary initial states \cite{beige,kastoryano,shankar,popkov} and can also be exploited for dissipative quantum computing \cite{verstraete} and quantum memories \cite{pastawski}. Dissipative state engineering processes are forceful, since, due to the dissipative nature of the process, the system is directed towards its nonequilibrium steady state (NESS)  regardless  the initial state of the system and the presence of perturbations \cite{popkov}. 
The system-environment dynamics can be described as  coherent Hamiltonian and dissipative open-system evolution \cite{Buca}. Here, we assume that the dynamic is purely dissipative; $i.e.$ the Hamiltonian of the system of our interest is vanished ( $H_{S} = 0$)  \cite{barreiro}. For years, noise induced by the environment has been considered as the enemy of quantum technologies. This is because very often the coupling between a quantum system and its environment leads to the quick disappearance of quantum properties, coherences, and entanglement. This perspective has been changed once scientist   established that suitable manipulation of an artificial environment (quantum reservoir engineering) would allow one to direct the open system towards, $e.g.$, a maximally entangled state \cite{barreiro,verstraete,barreiro}, hence changing the perspective of the environment as an enemy. The authors in  \cite{barreiro} simulated ion traps to demonstrate this engineering by dissipatively preparing a Bell state in a $N+1$ ion system. For example for $2+1$ ions,  an  initially fully mixed state is pumped into a given Bell state $e. g.$ $\ket{\psi^{-}}=1/\sqrt{2} \left(\ket{01}-\ket{10}\right)$.  Similarly, with $4+1$ ions, they  similarly  dissipatively prepared a $4-$qubit GHZ-state $\left(1/\sqrt{16}\left(\ket{0000}+\ket{1111}\right)\right)$. Afterward,  in \cite{verstraete}, the authors illustrated also this engineering by dissipatively preparing a Bell state in a $2+1$ ion system using  IBM Q \cite{ibm} experience processors publicly available and remotely accessible online.

In this paper, we prepare  four different Bell-state , $\ket{\phi}^{\pm}$  and $\ket{\psi}^{\pm}$. Additionally, we prepare  the GHZ state,    by applying necessarily strong dissipative couplings using  the QASM simulator \cite{qiskit}, and free web based interface, IBM Quantum Experience (IBM QE) \cite{ibm},  to show how   noises  affect  the   state's preparation as well as the population of the targeted Bell's and GHZ states.   This  dynamics is frequently described  as continuous in time, described by many-body Lindblad master equations, cf. $e.g.$ \cite{gardiner}. The targeted  entangled  state is  required to be an eigenstate with respect to the quantum jump operators. 
\section{Entanglement through Dissipation dynamics}
We discuss in this section, the engineering of the  interaction of a quantum system to its environment to achieve a preparation of this system to  a desired maximally entangled  state of $N$ qubits \cite{zoller}.
The dynamics of an open quantum system $S$ coupled
to an environment $E$ can be described by the unitary
transformation $\rho _{SE} \longrightarrow  U \rho_{SE} U^{\dagger}$ , where $\rho_{SE}$ represents the density operator of the joint state of system and its environment. Then, the reduced density operator of the system only will evolve as $\rho_{S} = Tr_{E}\left (U \rho_{SE} U^{\dagger}\right) $. The time evolution of the system can also be described by a completely positive Kraus map \cite{PhysRev.121.920},

\begin{equation}\label{eq1}
\rho_{S} \longrightarrow \mathcal{E}(\rho_{S}) = \sum^{K}_{k=1} E_{k} \rho_{SE} E^{\dagger}_{k}    
\end{equation}

where $\rho_S$ is the reduced density operator of the system, and $E_k$ are the so-called Kraus operators and satisfy the condition $\sum^{K}_{k=1} E_{k}^{\dagger} E_{k} = \mathbb{1}$ \cite{nielsen}. Whereas the structures of unitary maps on the system are attained for a single Kraus operator $K = 1$,   the dissipative dynamics  corresponds to multiple Kraus operators $K > 1$, and in this case the  evolution is in general non-unitary. Control of both coherent and dissipative dynamics is then accomplished by finding equivalent sequences of maps (\ref{eq1}) identified by sets of operation elements $\{E_{k} \}$ and engineering these sequences in the laboratory. In particular, for the example of dissipative quantum state preparation, pumping to an entangled state $\ket{\psi_{i}}$ reduces to implementing appropriate sequences of dissipative maps. These maps are chosen to direct the system to the desired target state irrespective of its initial state. The resulting dynamics have then the pure state $\ket{\psi_{i}}$ , $\rho_{S} \longrightarrow \ket{\psi_{i}}\bra{\psi_{i}}$. For $K=2$ , equation (\ref{eq1})  will be described, then,  as  the dissipative map pumping into the $\pm 1$ eigenspace  of the stabilizer of the target state,

\begin{equation}\label{eq2}
\rho_{S} \longrightarrow \mathcal{E}(\rho_{S}) = E_{1} \rho_{S} E_{1}^{\dagger} +E_{2} \rho_{S} E_{2}^{\dagger}
\end{equation}

 For dissipative map that pumps the system  into the $+1$ eigenspace of $\sigma^{z}_{1} \otimes \sigma^{z}_{2} \otimes ....\otimes \sigma^{z}_{N}$ (where $N$ represents the number of qubits in the system and $\sigma^{z}_{N}$ is the $z$- type Pauli operator acting on the $N^{th}$ qubit), $E_{1} = \sqrt{p} \mathbb{1} \otimes \sigma_{1}^{x} \frac{1}{2} \left( \mathbb{1} - A \right)$, and  $E_{2} = \frac{1}{2} \left(\mathbb{1} + A\right) + \sqrt{1-p}  \frac{1}{2} \left( \mathbb{1} - A\right)$, where $A = \sigma^{z}_{1} \otimes \sigma^{z}_{2} \otimes ....\otimes \sigma^{z}_{N} $  .  
 
 The map’s action as a uni-directional pumping process can be seen as follows: since the operation element $E_1$ contains the projector $\frac{1}{2} \left(\mathbb{1}-A \right)$ into $-1$ eigenspace of $A$ , the spin flip $\sigma^{x}_{1}$ can then convert $-1$ into $+1$ eigenstates of $A$. In contrast, the $+1$ eigenspace of $A$ is left invariant. The dissipative (cooling) dynamics are determined by the probability of cooling from the $-1$ into the $+1$
stabilizer eigenspaces, which can be directly controlled by varying the parameter $p$. For cooling with unit probability $(p = 1)$, the  intial mixed state  reach the target pure output state after only one cooling cycle. 
 
 The  map for cooling into the $+ 1$ eigenspace of $\sigma^{x}_{1} \otimes \sigma^{x}_{2} \otimes ....\otimes \sigma^{x}_{N}    $ is achieved  by interchanging the roles of $\sigma^{x}$ and $\sigma^{z}$. The ground state(s) is/are therefore given by the simultaneous eigenstate(s) of all stabilizers with eigenvalues $+1$.  
The Markovian limit of these open-system dynamics, in equation (\ref{eq2}),  can be written as a master Lindblad  equation \cite{lindblad,gorini},

\begin{equation}\label{eq3}
\frac{d \rho_{S}}{dt}  =   -i \left[H_{S},\rho_{S}\right] +\\   \sum_{k} \gamma_{k}  \left( c_{k} \rho_{S}c_{k}^{\dagger}
- \frac{1}{2} c_{k}^{\dagger}c_{k}\rho_{S} 
   -\rho_{S}\frac{1}{2}c_{k}^{\dagger}c_{k} \right)
\end{equation}

Here, $H_S$ is the system Hamiltonian, $c_{k}$ are  Lindblad (or quantum jump) operators, and  $\gamma_{k}$ are  dissipative rates reflecting the strength of the continuous interaction between the system and its environment. For the master equation, essential conditions to complete such dissipative dynamics which contracts to a pure state is given by the conditions 
$H_{S} \ket{\psi} = E \ket{\psi}$ and $\forall k$ $c_{k} \ket{\psi} = 0$. As the latest condition illustrates, $\ket{\psi}$ is a system decoupled from the environment, which in quantum optics is called \quotes{ dark state} \cite{barreiro}. For the purely dissipative maps discussed in this article, $H_{S} = 0$. In the limit $p << 1$ in Eq. (\ref{eq2}), after repeating the action of this map, then, the  initial mixed state  will reach the target pure output state after multiple cooling cycles. The quantum jump operators in Eq. (\ref{eq3}) is described mathematically as, 

\begin{equation}\label{eq4}
    c_{1}= \frac{1}{2} \sigma_{1}^{x} \left(1- A\right) \quad\text{and}\quad    c_{2}= \frac{1}{2} \sigma_{1}^{z} \left(1- B\right)
\end{equation}
where, $B = \sigma^{x}_{1} \otimes \sigma^{x}_{2} \otimes ....\otimes \sigma^{x}_{N} $
\subsection{ Cooling to Bell's state}
\begin{figure}[ht]
     \centering
     \begin{subfigure}{0.5\textwidth}
         \centering
         \includegraphics[width=\textwidth]{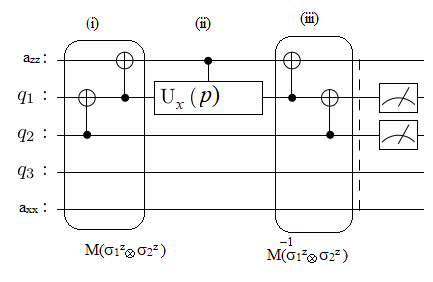}
         \caption{.}
         \label{.}
     \end{subfigure}
    
     \begin{subfigure}{0.5\textwidth}
         \centering
         \includegraphics[width=\textwidth]{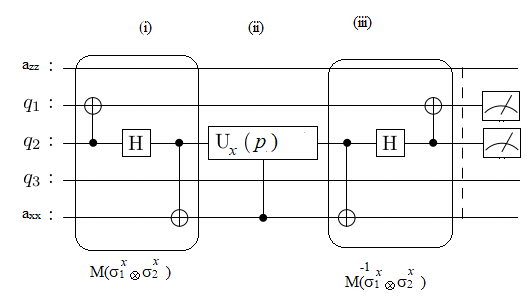}
         \caption{,}
         \label{,}
     \end{subfigure}

        \caption{The quantum circuits performing the cooling (pumping) of an arbitrary initial mixed   state  (not shown) of qubits $q_{1}$ and $q_{2}$ into $\pm 1$ eigenspace of   (a) $\sigma^{z}_{1} \otimes \sigma^{z}_{2}$, $\mathcal{E}_{z_{1}z_{2}}(\rho_{S})$ map,   and  (b) $\sigma^{x}_{1} \otimes \sigma^{x}_{2}$, $\mathcal{E}_{x_{1}x_{2}}(\rho_{S})$ map  . The circuits were run on QASM simulator and also IBM Q processor,  with and without errors mitigation. Qubits $a_{xx}$ and $a_{zz}$ represent the environment ancillae for the two maps.}
        \label{fig:fig1}
\end{figure}

In this section, we investigate   the dissipative preparation  of the   Kraus's map engineering for the simplest  example of “cooling” a quntum system of an arbitrary initial mixed state of two qubits into a Bell state. The Hilbert space of the two qubits is spanned by the four Bell states described  as $\ket{\phi^{\pm}} =\frac{1}{\sqrt{2}}\left( \ket{00} \pm \ket{11} \right)$ and $\ket{\psi^{\pm}} =\frac{1}{\sqrt{2}}\left( \ket{01} \pm \ket{10} \right)$. Here, $\ket{0}$ and $\ket{1}$ represent the computational basis of each qubit, and we use the short-hand
notation $\ket{00} = \ket{0}_{1} \ket{0}_{2}$, we notice that these states are mutual eigenstates of the two commuting stabilizer operators  $\sigma^{z}_{1} \otimes \sigma^{z}_{2}$ and $\sigma^{x}_{1} \otimes \sigma^{x}_{2}$ with eigenvalues $\pm 1$. For example, the Bell state $\ket{\phi^{+}}$ is the  stabilized state  by the two stabilizer operators $\sigma^{z}_{1} \otimes \sigma^{z}_{2}$ and $\sigma^{x}_{1} \otimes \sigma^{x}_{2}$, as it is the only two qubit state presence an eigenstate of eigenvalue $+1$ of these two commuting operators, $i. e.$ $\sigma^{z}_{1} \otimes \sigma^{z}_{2}\ket{\phi^{+}} = \ket{\phi^{+}} $ and $\sigma^{x}_{1} \otimes \sigma^{x}_{2}\ket{\phi^{+}} = \ket{\phi^{+}} $.  The cooling concept can be realized by dissipative dynamics which cool (pump) the arbitrary mixed state into one of Bell’s state, for example, $\rho_{S} \Longrightarrow \ket{\phi^{+}} \bra{\phi^{+}}$, by implementing  two dissipative maps, under which the two qubits are irreversibly transferred from the $-1$ into the $+1$ eigenspaces of $\sigma^{z}_{1} \otimes \sigma^{z}_{2}$ and $\sigma^{x}_{1} \otimes \sigma^{x}_{2}$. The dissipative maps are engineered with the assistance of an ancilla '' artificial environment'' qubit \cite{barreiro,Lloyd, Dur} and a quantum circuit of dissipative operations and a quantum circuit of  dissipative operations as shown in Fig. (\ref{fig:fig1}).

Specially, we consider the two $p$-parameterised families as in equation (\ref{eq2})

\begin{equation}\label{eq5}
   \mathcal{E}_{zz} \left(\rho_{S}\right) =  E_{1z} \rho_{S} E_{1z}^{\dagger} +E_{2z} \rho_{S} E_{2z}^{\dagger}
 \end{equation}
where,  $E_{1z} = \sqrt{p} \mathbb{1}_{2} \otimes \sigma_{1}^{x} \frac{1}{2} \left( \mathbb{1} -\sigma^{z}_{1} \otimes \sigma^{z}_{2} \right)$, and  $E_{2z} = \frac{1}{2} \left(\mathbb{1} + \sigma^{z}_{1} \otimes \sigma^{z}_{2}\right) + \sqrt{1-p}  \frac{1}{2} \left( \mathbb{1} -\sigma^{z}_{1} \otimes \sigma^{z}_{2}\right)$. Where, $\mathcal{E}_{xx} \left(\rho_{S}\right)$ has the same form of Eq. (\ref{eq5}), but we exchange the role of  $\sigma_{x}$ and  $\sigma_{z}$ in all terms in Eq (\ref{eq5}). By varying the parameter $0\le p \le 1$ in $E_{1z}$ and $E_{2z}$, we simulate various forms of open quantum system dynamics. For $ p << 1$ , the redundant application of $\mathcal{E}_{zz} $ and  $\mathcal{E}_{xx} $ generates a master equation of Lindblad form with jump operator  as in Eq. (\ref{eq4}). For Bell's state cooling, $A= \sigma^{z}_{1} \otimes \sigma^{z}_{2}$, and $B= \sigma^{z}_{1} \otimes \sigma^{z}_{2}$ \cite{guillermo}. For cooling with unit probability
$(p = 1)$, the two qubits reach the target Bell state - irrespective of their initial state - after only one cooling cycle, $i.e$., by a single application of each of the two maps, $i.e.$ $\mathcal{E}_{zz}$ and $\mathcal{E}_{xx}$ to generate for example $\ket{\phi^{+}}$.

\begin{figure}[ht]
  \includegraphics[width=1.0\linewidth]{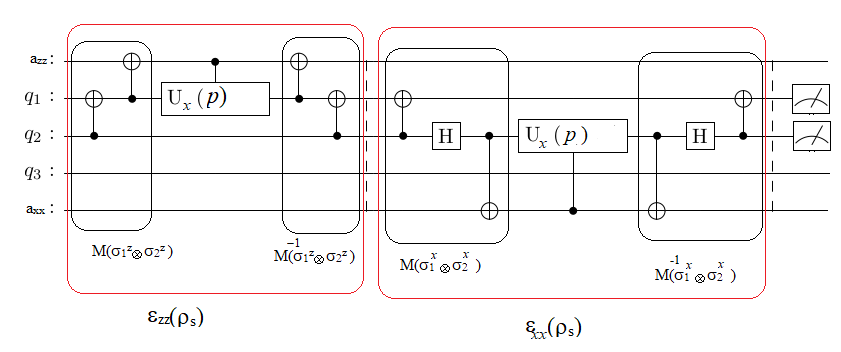}
  \caption{The quantum circuit performing the cooling (pumping) of an arbitrary initial mixed   state  (not shown) of qubits $q_{1}$ and $q_{2}$ into $\pm 1$ eigenspace of    $\sigma^{z}_{1} \otimes \sigma^{z}_{2}$   and   $\sigma^{x}_{1} \otimes \sigma^{x}_{2}$, $\mathcal{E}_{x_{1}x_{2}} \left(\mathcal{E}_{z_{1}z_{2}  }\left(\rho(S)\right)\right)$ map  . The circuits were run on QASM simulator and also IBM Q processor,  with and without errors mitigation. Qubits $a_{xx}$ and $a_{zz}$ represent the environment ancillae for the two map.}
  \label{fig:fig2}
\end{figure}

The  circuits  for the execution  of the Bell's state cooling originally suggested  by   ref.\cite{barreiro}, their circuits are composed of gates that are appropriate with trapped-ions platform. However, the authors in  ref.\cite{guillermo} provided   circuits that follow the same  basic functioning principles, but have been designed specifically keeping in mind the characteristics of the IBM Q Experience device \cite{ibm}. In ref \cite{barreiro,guillermo}, authors  cool the arbitrary initial mixed state into $\ket{\psi^{-}}$. In this article,  we cool the initial state into  four different Bell's states, $\ket{\psi^{-}}$, $\ket{\psi^{+}}$, $\ket{\phi^{-}}$, and $\ket{\phi^{+}}$,  by controlling  the environment state for each targeted output state.  We simulate the circuits in Fig. (\ref{fig:fig1}) and Fig. (\ref{fig:fig2}) using QASM simulator \cite{qiskit} and further using real IBM quantum device  with and without error mitigation's technique \cite{ibm}, to study the effect of the noise on  preparation of the initial mixed state of the two qubits as well as the population of the target state. 

To prepare the  two-qubit system initially in the maximally mixed state $\rho_{s} = \mathbb{1}_{4}/4$, where $\mathbb{1}_{4}$ is the $4 \times 4$ identity matrix. Technically, this could be done by entangling the system with other ancillary qubits, but that would require two extra qubits in our simulation. Instead, we can create a proper statistical mixture. This means that we can obtain $\rho_{S}$ by mixing four initially pure states, $e.g.$ the two-qubit computational basis states\cite{guillermo}. Accordingly, we prepare the initial maximally mixed state $\rho_{S}$ by mixing four initially pure states, $e.g.$ the two-qubit computational basis state. For each initial state of the qubits $\ket{00}$,  $\ket{01}$, $\ket{10}$, and $\ket{11}$, we apply the three channels $\mathcal{E}_{zz}$, $\mathcal{E}_{xx}$, and 
$\mathcal{E}_{zz}\left(  \mathcal{E}_{xx}\right)$ for different values of $p \in \left[0,1\right]$. To finally simulate the effect of the different channels on the maximally mixed state, we average all the results over the four initial states. Plot the output target state population as a function of the channel efficiency  $p$.

The pumping (cooling) circuits proposed in ref. \cite{barreiro} and \cite{gardiner} implementing  the two dissipative maps  are composed of three unitary operators $(i)$, $(ii)$, and $(iii)$ as shown in Fig (\ref{fig:fig1}). First, the information about the state of the system  whether it is  in the $+1$ or  $-1$ eigenspaces of the $\sigma^{z}_{1} \otimes  \sigma^{z}_{2}$ stabilizer operator, for example, is mapped into an ancilla $\ket{0}$ or$\ket{1}$ (originally $\ket{0}$) $(i)$. Second, the state of the system is changed depending on the state of the ancilla  using  controlled gate $U_{x} (p) (ii)$, converts $\pm1$ into $\mp1$ eigenstates by flipping the state of the first qubit with probability $p$, where,
\begin{equation}\label{eq6}
   U_{x} (p)  = \ket{0}\bra{0} \otimes \mathbb{1} + \ket{1}\bra{1} \otimes \sigma_{x}
   \end{equation}
Where $U_{x} (p)  = exp \left( i \theta \sigma_{x}\right)$ and $p = sin^{2}$ $\theta$. Third, the mapping circuit is reversed $(iii)$. For $p=1$ $e.g.$ deterministic cooling (pumping), the circuit simplifies as the inverse mapping to (step ($iii$))   is not required, and the circuit simplifies to only  two-step process \cite{Schindler}. By the end of the last step the system has been pumped into the desired  Bell's state. The circuit in \cite{barreiro} contains four steps instead of three, whereas the fourth step is designated for the ancilla reset (pumping) to use it again for the following pumping cycle. However, in \cite{gardiner} as well as this article, the fourth step has been neglected   since the IBM Q Experience devices are not furnished with the reset operation,  so, we should prepare  an ancilla for every pumping cycle. 

The second map for cooling into the $\pm 1$ eigenspace of $\sigma^{x}_{1} \otimes  \sigma^{x}_{2}$ is obtained from interchanging the roles of $\sigma^{x}$ and $\sigma^{z} $ above. We apply the circuits  in Fig. (\ref{fig:fig1})  for $\mathcal{E}_{zz}\left(\rho_{S}\right)$, $\mathcal{E}_{xx}\left(\rho_{S}\right)$, and Fig (\ref{fig:fig2}) for $\mathcal{E}_{xx}\left(\mathcal{E}_{zz}\left(\rho_{S}\right)\right)$ on the arbitrary mixed state that we discussed above to pump  this initial state into one of the  four Bell's state. We run the quantum circuits  using both, QASM simulator \cite{ibm}, and furthermore, using  one of the  IBM quantum processors  that consists  4-qubits, as those are allowed  to be used for free  publicly. Through controlling (varying) the environment state  $(a_{zz}$ and $a_{xx})$ in Fig (\ref{fig:fig2}),   the  system  will be pumped (cooled)  to the desired Bell's state as shown in Table (\ref{tab:table1}). For example, if  $a_{zz}$ = $a_{xx}$ = $\ket{0}\bra{0}$, then the initial mixed state $\rho_{S}$ will be pumped to  $+1$ eigenspace of the stabilizer operators $\sigma^{z}_{1} \otimes \sigma^{z}_{2}$  and $\sigma^{x}_{1} \otimes \sigma^{x}_{2}$ $i.e.$ $\rho_{S} \longrightarrow \ket{\phi^{+}}\bra{\phi^{+}}$. Furthermore, to pump the initial state to  $+1$ of the stabilizer operator $\sigma^{z}_{1} \otimes \sigma^{z}_{2}$ and $-1$ eigenspace of the stabilizer operator $\sigma^{x}_{1} \otimes \sigma^{x}_{2}$ $i.e.$ $\rho_{S} \longrightarrow \ket{\phi^{-}}\bra{\phi^{-}}$, then  we   adjust the state of $a_{xx}$ to $\ket{1}\bra{1}$. However, if we vary the state of   $a_{zz}$ to $\ket{1}\bra{1}$, then the initial state will be targeted  to $\ket{\psi^{+}}\bra{\psi^{+}}$ ( $-1$  eigenspace of the stabilizer operator $\sigma^{z}_{1} \otimes \sigma^{z}_{2}$ and $+1$ eigenspace of the stabilizer operator $\sigma^{z}_{1} \otimes \sigma^{z}_{2}$). Finally, to pump the system to $\ket{\psi^{-}}\bra{\psi^{-}}$ ( $-1$  eigenspace of the stabilizer operators $\sigma^{z}_{1} \otimes \sigma^{z}_{2}$ and  operator $\sigma^{z}_{1} \otimes \sigma^{z}_{2}$), then, $a_{zz}$ = $a_{xx}$ = $\ket{1}\bra{1}$.
\begin{table}[h!]
  \begin{center}
    \caption{Table of possible quantum state of the environment, $a_{z_{1}z_{2}}$ (first column) and   $a_{x_{1}x_{2}}$ (the second column) ,as shown in Fig. (\ref{fig:fig1}) and Fig. (\ref{fig:fig2}), and its  associated  targeted Bell-state (third column)}
    \label{tab:table1}
    \begin{tabular}{|l|c|r| } 
    \hline
     $a_{z_{1}z_{2}}$ & $a_{x_{1}x_{2}}$ & $\rho_{out}(S)$\\
      \hline
      $\ket{0}\bra{0}$ & $\ket{0}\bra{0}$ &$ \ket{\phi^{+}}\bra{\phi^{+}}$\\
      \hline
      $\ket{0}\bra{0}$ & $\ket{1}\bra{1}$ & $\ket{\phi^{-}}\bra{\phi^{-}}$\\
      \hline
      $\ket{1}\bra{1}$ & $\ket{0}\bra{0}$ &$\ket{\psi^{+}}\bra{\psi^{+}}$ \\
      \hline
      $\ket{1}\bra{1}$ & $\ket{1}\bra{1}$ & $\ket{\psi^{-}}\bra{\psi^{-}}$ \\
      \hline

    \end{tabular}
  \end{center}
\end{table}

\begin{figure}[ht]
\begin{subfigure}{0.3\textwidth}
  \centering
  \includegraphics[width=\textwidth]{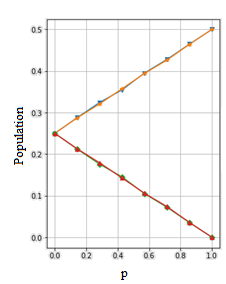}
  \caption{.}
\end{subfigure}%
\begin{subfigure}{0.3\textwidth}
  \centering
  \includegraphics[width=\textwidth]{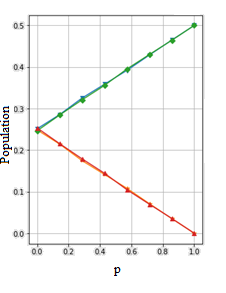}
    \caption{,}

\end{subfigure}
\begin{subfigure}{0.3\textwidth}
  \centering
  \includegraphics[width=\textwidth]{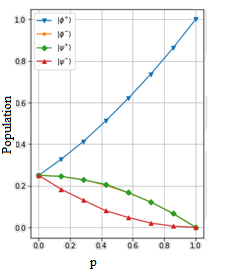}
    \caption{.}
\end{subfigure}

  \caption{Simulation of the Markovian Lindblad master equationn using QASM simulator. Evolution of the Bell-state populations for $a_{z_{1}z_{2}} = a_{x_{1}x_{2}} = 0$  for  $\sigma^{z}_{1} \otimes \sigma^{z}_{2}$  cooling (a),$\sigma^{x}_{1} \otimes \sigma^{x}_{2}$ cooling(b) and their consecutive application (c).The cooling processes are performed to an initial maximally mixed state.   Error bars, not shown, are smaller than $2\%$}
  \label{fig:fig3}
\end{figure}

\begin{figure}[ht]
  \includegraphics[width=0.7\linewidth]{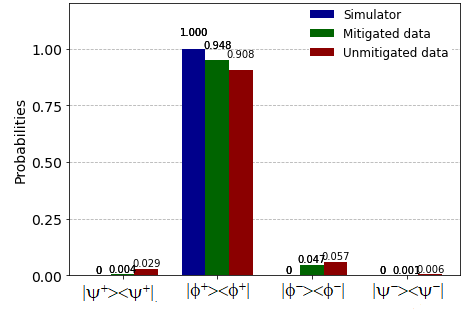}
  \caption{Probability of the population of the targeted Bell's simulation resulting by QASM simulation, and IBM Q processors (without and with errors mitigation) for the case $a_{zz} = a{xx}=0$}
  \label{fig:fig4}
\end{figure}

\begin{figure}[ht]
\begin{subfigure}{0.3\textwidth}
  \centering
  \includegraphics[width=\textwidth]{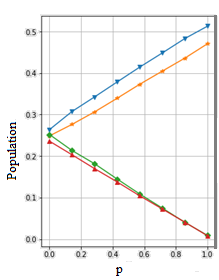}
  \caption{.}
\end{subfigure}%
\begin{subfigure}{0.3\textwidth}
  \centering
  \includegraphics[width=\textwidth]{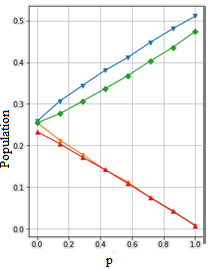}
    \caption{.}

\end{subfigure}
\begin{subfigure}{0.3\textwidth}
  \centering
  \includegraphics[width=\textwidth]{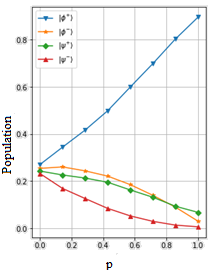}
    \caption{.}
\end{subfigure}

  \caption{Simulation of the Markovian Lindblad master equation using IBM processor without errors mitigation. Evolution of the Bell-state populations for $a_{z_{1}z_{2}} = a_{x_{1}x_{2}} = 0$  for  $\sigma^{z}_{1} \otimes \sigma^{z}_{2}$  cooling (a),$\sigma^{x}_{1} \otimes \sigma^{x}_{2}$ cooling(b) and their consecutive application (c).The cooling processes are performed to an initial maximally mixed state.   Error bars, not shown, are smaller than $2\%$}
    \label{fig:fig5}

\end{figure}
\begin{figure}[ht]
\begin{subfigure}{0.3\textwidth}
  \centering
  \includegraphics[width=\textwidth]{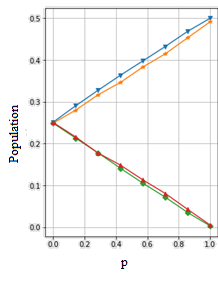}
  \caption{.}
\end{subfigure}%
\begin{subfigure}{0.3\textwidth}
  \centering
  \includegraphics[width=\textwidth]{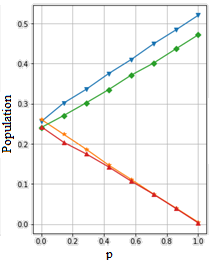}
    \caption{.}
\end{subfigure}
\begin{subfigure}{0.3\textwidth}
  \centering
  \includegraphics[width=\textwidth]{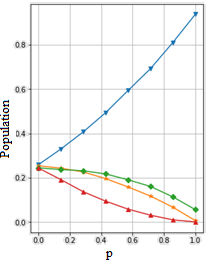}
    \caption{.}
\end{subfigure}
  \caption{Simulation of the Markovian Lindblad master equation using IBM processor with errors mitigation. Evolution of the Bell-state populations for $a_{z_{1}z_{2}} = a_{x_{1}x_{2}} = 0$  for  $\sigma^{z}_{1} \otimes \sigma^{z}_{2}$  cooling (a),$\sigma^{x}_{1} \otimes \sigma^{x}_{2}$ cooling(b) and their consecutive application (c).The cooling processes are performed to an initial maximally mixed state.   Error bars, not shown, are smaller than $2\%$}
    \label{fig:fig6}
\end{figure}

We illustrate  the population of the Bell's state versus the pumping probability $p$,  using  QASM simulator Fig. (\ref{fig:fig3}) and also the  noisy IBM Q quantum processor, without and with errors mitigation,  Fig. (\ref{fig:fig4}), Fig.(\ref{fig:fig5}),respectively, when  $a_{zz} = a_{xx} = \ket{0}\bra{0}$. As expected, the initial mixed state has been pumped into $\ket{\phi^{+}} \bra{\phi^{+}}$.  One of the primary concerns of this study is to show how   noises in the IBM Q processors affect  the   state's preparation as well as the population of the targeted Bell's state.   The effect of noise that occurs throughout the  computation will be quite complex in general, as one would have to consider how each gate transforms the effect of each error. A simpler form of noise is that occurring during  state's preparation and final measurement. QISKIT platform \cite{qiskit} has a feature that can mitigate those types of errors. In addition, we notice that QASM simulator  prepares  the two qubits in a perfect maximally mixed  state, $e.g$ $ \frac{1}{4} \left(\ket{00}\bra{00} + \ket{01}\bra{01} + ... +\bra{11}\bra{11} \right)$. Whereas,  IBM Q quantum processors fail to prepare such a state due to using noisy gates. Moreover,  we observe in Fig.(\ref{fig:fig3}), the simulator case,  the target state population is preserved under the frequent execution of additional  pumping   cycles and reaches up to (100$\%$), (90.7$\%$) for unmitigated case (Fig. (\ref{fig:fig4})), and (94.8$\%$) for the mitigated case (Fig. (\ref{fig:fig5})), as shown also in Fig. (\ref{fig:fig4}).

\begin{figure}[ht]
\begin{subfigure}{0.3\textwidth}
  \centering
  \includegraphics[width=\textwidth]{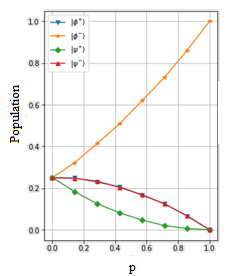}
  \caption{.}
\end{subfigure}%
\begin{subfigure}{0.3\textwidth}
  \centering
  \includegraphics[width=\textwidth]{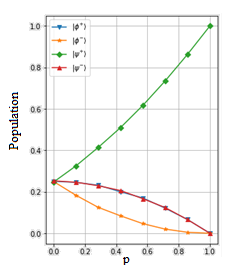}
    \caption{.}

\end{subfigure}
\begin{subfigure}{0.3\textwidth}
  \centering
  \includegraphics[width=\textwidth]{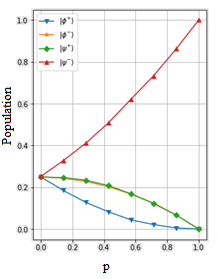}
    \caption{.}
\end{subfigure}

\caption{Simulation of the Markovian Lindblad master equation using QASM simulator  only for  the consecutive application of two the maps $\mathcal{E}_{xx} \left(\mathcal{E}_{zz}\right)$ for $a_{zz} = 0$ and $a_{xx} = 1$ (a), $a_{zz} = 0$ and $a_{xx} = 1$ (b), and $a_{zz} = a_{xx} = 1$ (c)}
 \label{fig:fig7}
\end{figure}

\begin{figure}[ht]
\begin{subfigure}{0.3\textwidth}
  \centering
  \includegraphics[width=\textwidth]{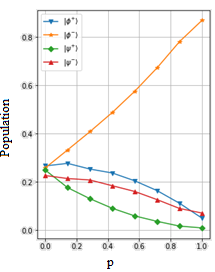}
  \caption{.}
\end{subfigure}%
\begin{subfigure}{0.3\textwidth}
  \centering
  \includegraphics[width=\textwidth]{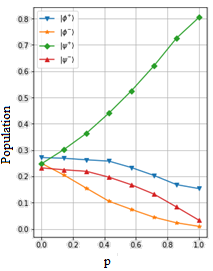}
    \caption{.}

\end{subfigure}
\begin{subfigure}{0.3\textwidth}
  \centering
  \includegraphics[width=\textwidth]{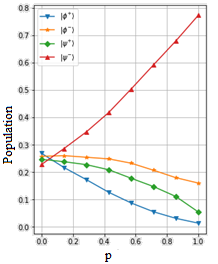}
    \caption{.}
\end{subfigure}

\caption{Simulation of the Markovian Lindblad master equation using IBM Q processor without errors mitigation simulator  only for  the consecutive application of two the maps $\mathcal{E}_{xx} \left(\mathcal{E}_{zz}\right)$ for $a_{zz} = 0$ and $a_{xx} = 1$ (a), $a_{zz} = 0$ and $a_{xx} = 1$ (b), and $a_{zz} = a_{xx} = 1$ (c)}
\label{fig:fig8}
\end{figure}

By  varying the state of the environment,  $a_{zz}$ and  $a_{xx}$, as shown in table (\ref{tab:table1}), drives  the initial input state  into other Bell's state. In particular,  Fig.(\ref{fig:fig6}.a) (simulator case), Fig.(\ref{fig:fig7}.a) (IBM processor without errors mitigation), an Fig.(\ref{fig:fig8}.a) (IBM processor with errors mitigation)   show  cooling into $\ket{\phi^{-}}$ for the case $a_{z_{1}z_{2}} = 0$ and  $a_{x_{1}x_{2}} = 1$ as expected in table (\ref{tab:table1}). Fig.(\ref{fig:fig6}.b) (simulator case), Fig.(\ref{fig:fig7}.b) (IBM processor without errors mitigation), an Fig.(\ref{fig:fig8}.c) (IBM processor with errors mitigation)   show  cooling into $\ket{\psi^{+}}$ for the case $a_{z_{1}z_{2}} = 1$ and  $a_{x_{1}x_{2}} = 0$. Finally, for  $a_{z_{1}z_{2}} = 1 a_{x_{1}x_{2}} = 1$, the initial input state for the system is cooling to $\ket{\psi^{-}}$ as presented in Fig.(\ref{fig:fig6}.c) (simulator case), Fig.(\ref{fig:fig7}.c) (IBM processor without errors mitigation), and Fig.(\ref{fig:fig8}.c) . Here, we present the graphs only for $\mathcal{E}_{xx} \left (\mathcal{E}_{zz}\right)$, $i.e.$ the targeted Bell's state after performing the dissipate maps, $\mathcal{E}_{xx}$ and $ \mathcal{E}_{zz}$.

\begin{figure}[ht]
\begin{subfigure}{0.3\textwidth}
  \centering
  \includegraphics[width=\textwidth]{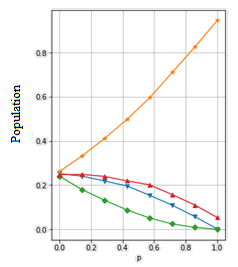}
  \caption{.}
\end{subfigure}%
\begin{subfigure}{0.3\textwidth}
  \centering
  \includegraphics[width=\textwidth]{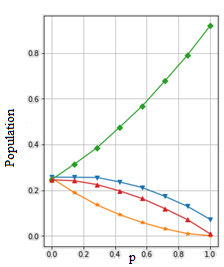}
    \caption{.}

\end{subfigure}
\begin{subfigure}{0.3\textwidth}
  \centering
  \includegraphics[width=\textwidth]{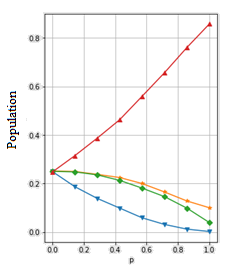}
    \caption{.}
\end{subfigure}

\caption{Simulation of the Markovian Lindblad master equation using IBM Q processor with errors mitigation simulator  only for  the consecutive application of two the maps $\mathcal{E}_{xx} \left(\mathcal{E}_{zz}\right)$ for $a_{zz} = 0$ and $a_{xx} = 1$ (a), $a_{zz} = 0$ and $a_{xx} = 1$ (b), and $a_{zz} = a_{xx} = 1$ (c)}
\label{fig:fig9}
\end{figure}

\subsection{ Cooling to GHZ state}
\begin{figure}[ht]
  \includegraphics[width=1.0\linewidth]
  {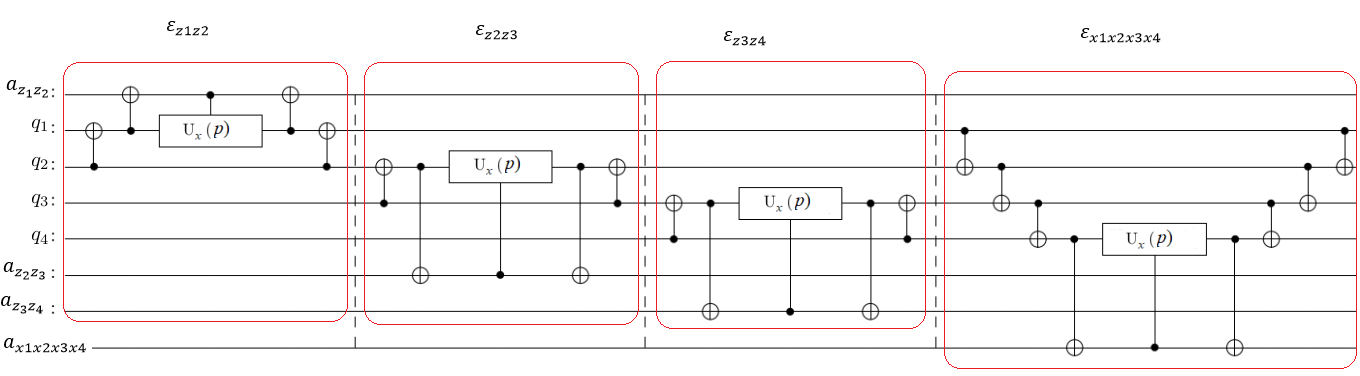}
  \caption{The quantum circuit performing the cooling (pumping) of an arbitrary initial mixed state (not shown) of
qubits $q_1$, $q_2$, $q_3$, and $q_4$ into $\pm1$ eigenspace of $\sigma^{z}_{1} \otimes \sigma^{z}_{2}$, $\sigma^{z}_{2} \otimes \sigma^{z}_{3}$, $\sigma^{z}_{3} \otimes \sigma^{z}_{4}$ and $\sigma^{x}_{1} \otimes \sigma^{x}_{2} \otimes \sigma^{x}_{3} \otimes \sigma^{x}_{4} $. The circuit were run on QASM simulator and also IBM Q processor, with and without errors mitigation. Qubits $a_{z_{1}z{2}}$, $a_{z_{2}z{3}}$, $a_{z_{3}z{4}}$, and  $a_{x_{1}x{2}x{3}x{4}}$ 
represent the environment ancillae for the four maps.
}
 \label{fig:fig10}
\end{figure}

Monitoring  the system-environment coupling and also the state of the environment, as presented  by Bell-state cooling in the previous section, can  be undoubtedly  extended to larger $n$-qubit open quantum systems. In this section, we  monitor  the system-environment interaction  dissipatively  of a four-qubit Greenberger-Horne-Zeilinger (GHZ) state $\frac{1}{\sqrt{2}}\left(\ket{0000} + \ket{1111} \right)$. This state is exclusively considered as the simultaneous eigenstate of the four stabilizers $\sigma^{z}_{1} \otimes \sigma^{z}_{2}$,  $\sigma^{z}_{2} \otimes \sigma^{z}_{3}$, $\sigma^{z}_{3} \otimes \sigma^{z}_{4}$, and $\sigma^{x}_{1} \otimes \sigma^{x}_{2} \otimes \sigma^{x}_{3} \otimes \sigma^{x}_{4}$, all with eigenvalue $+1$.  Thus, cooling an arbitrary initial mixed state of four qubits system $i.e.$ $\frac{1}{16} \left(\ket{0000}\bra{0000} + \ket{0111}\bra{0111} + ... + \ket{1111}\bra{1111}\right)$,  into the GHZ state can be recognized by four successive dissipative maps, $\mathcal{E}_{z_{1}z_{2}}$, $\mathcal{E}_{z_{2}z_{3}}$, $\mathcal{E}_{z_{3}z_{4}}$, and $\mathcal{E}_{x_{1}x_{2}x_{3}x_{4}}$, as shown in the quantum circuit in Fig. (\ref{fig:fig10}), each pumping the system into the $+1$ eigenspaces of the four stabilizers. For example, $\mathcal{E}_{z_{1}z_{2}}$ pumps the system into the eigenspace of $\sigma^{z}_{1} \otimes \sigma^{z}_{2}$ with $+1$ eigenvalue, similary for other operators ($ {stabilizers}$). Therefore, for cooling to GHZ state for example,  we carry out the same cooling dynamics procedure we present for Bell’s state cooling from the previous section. The maps, $\mathcal{E}_{z_{1}z_{2}}$, $\mathcal{E}_{z_{2}z_{3}}$, $\mathcal{E}_{z_{3}z_{4}}$ which displaced in Fig. (\ref{fig:fig10}) cool (pump)  an initial  mixed State of four qubits state into $\frac{1}{\sqrt{2}} \left(\ket{0000} \pm \ket{1111} \right)$, as they are exclusively   designated  as the simultaneous eigenstate of $\mathcal{E}_{z_{1}z_{2}}$, $\mathcal{E}_{z_{2}z_{3}}$, $\mathcal{E}_{z_{3}z_{4}}$ with $+1$ eigenvalue. The last step is to direct  the state into the GHZ state  using the map $\mathcal{E}_{x_{1}x_{2}x_{3}x_{4}}$,  the last part in Fig. (\ref{fig:fig10}.

Table (\ref{tab:table2}) shows the cooling of the mixed state into sixteen different entangled  states of four qubits, including the GHZ state by controlling the state of the environment $a_{z_{1}z{2}}$, $a_{z_{2}z{3}}$, $a_{z_{3}z{4}}$, and  $a_{x_{1}x_{2}x_{3}x_{4}}$.
\onecolumngrid

\begin{table}[ht]
  \begin{center}
    \caption{Table of possible quantum state of the environment, $a_{z_{1}z{2}}$, $a_{z_{2}z{3}}$, $a_{z_{3}z{4}}$, and  $a_{x_{1}x{2}x{3}x{4}}$ and its associated  entangled sate of four qubits}
    \label{tab:table2}
    \begin{tabular}{l|c|c|c|r} 
    \hline
     $a_{z_{1}z_{2}}$ &  $a_{z_{2}z_{3}}$ &  $a_{z_{3}z_{4}}$& $a_{x_{1}x_{2}x_{3}x_{4}}$ & $\rho_{out}(S)$ \\
      \hline
      $\ket{0}$ & $\ket{0}$ & $\ket{0}$ & $\ket{0}$ & $(1/\sqrt{2})\left(\ket{0000} + \ket{1111}\right)$\\
      \hline
      $\ket{1}$ & $\ket{0}$ & $\ket{0}$ & $\ket{0}$ & $(1/\sqrt{2})\left(\ket{1010} + \ket{0101}\right)$\\
      \hline
      $\ket{0}$ & $\ket{1}$ & $\ket{0}$ & $\ket{0}$ & $(1/\sqrt{2})\left(\ket{0010} + \ket{1101}\right)$\\
      \hline
       $\ket{0}$ & $\ket{0}$ & $\ket{1}$ & $\ket{0}$ & $(1/\sqrt{2})\left(\ket{0001} + \ket{1110}\right)$\\
       \hline
        $\ket{0}$ & $\ket{0}$ & $\ket{0}$ & $\ket{1}$ & $(1/\sqrt{2})\left(\ket{0000} - \ket{1111}\right)$\\
        \hline
         $\ket{1}$ & $\ket{1}$ & $\ket{0}$ & $\ket{0}$ & $(1/\sqrt{2})\left(\ket{0111} + \ket{1000}\right)$\\
         \hline
         $\ket{1}$ & $\ket{0}$ & $\ket{1}$ & $\ket{0}$ & $(1/\sqrt{2})\left(\ket{1011} + \ket{0100}\right)$\\
         \hline
         $\ket{1}$ & $\ket{0}$ & $\ket{1}$ & $\ket{1}$ & $(1/\sqrt{2})\left(\ket{1010} - \ket{0101}\right)$\\
         \hline
         $\ket{0}$ & $\ket{1}$ & $\ket{0}$ & $\ket{1}$ & $(1/\sqrt{2})\left(\ket{0010} - \ket{1101}\right)$\\
         \hline
         $\ket{0}$ & $\ket{0}$ & $\ket{1}$ & $\ket{1}$ & $(1/\sqrt{2})\left(\ket{0001} - \ket{1110}\right)$\\
         \hline
         $\ket{1}$ & $\ket{1}$ & $\ket{1}$ & $\ket{0}$ & $(1/\sqrt{2})\left(\ket{0110} + \ket{1001}\right)$\\
         \hline
         $\ket{0}$ & $\ket{1}$ & $\ket{1}$ & $\ket{1}$ & $(1/\sqrt{2})\left(\ket{0011} - \ket{1100}\right)$\\
         \hline
         $\ket{1}$ & $\ket{0}$ & $\ket{1}$ & $\ket{1}$ & $(1/\sqrt{2})\left(\ket{1011} - \ket{0100}\right)$\\
         \hline
         $\ket{1}$ & $\ket{1}$ & $\ket{0}$ & $\ket{1}$ & $(1/\sqrt{2})\left(\ket{0111} - \ket{1000}\right)$\\
         \hline
         $\ket{1}$ & $\ket{0}$ & $\ket{1}$ & $\ket{1}$ & $(1/\sqrt{2})\left(\ket{1011} - \ket{0100}\right)$\\
         \hline
         $\ket{1}$ & $\ket{1}$ & $\ket{1}$ & $\ket{1}$ & $(1/\sqrt{2})\left(\ket{0110} - \ket{1001}\right)$\\
      \hline
    \end{tabular}
  \end{center}
\end{table}

\begin{figure}[ht]
  \includegraphics[width=1.0\linewidth]{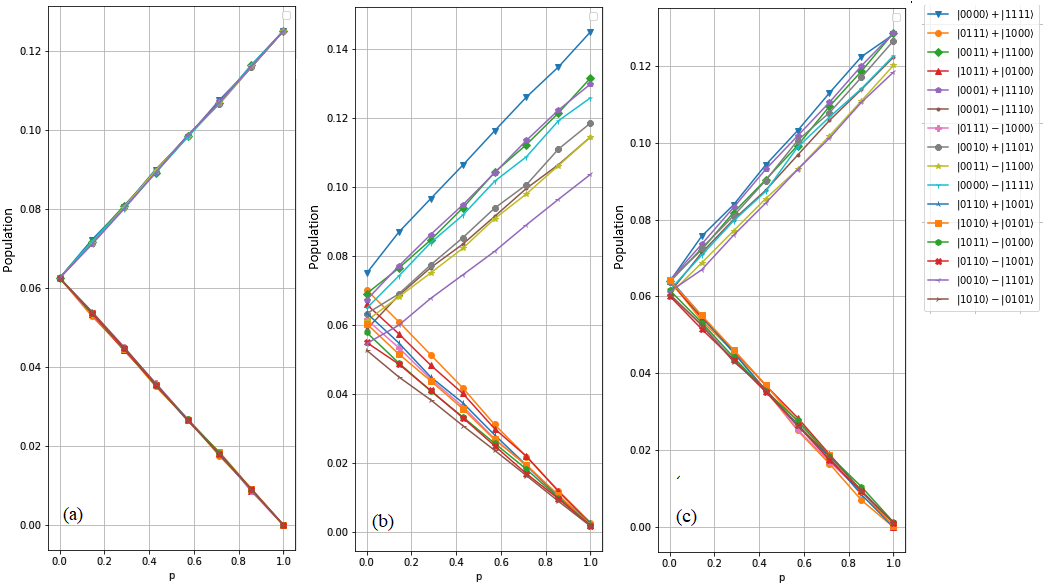}
  \caption{Simulation of the Markovian Lindblad master equation using QASM simulate (a), IBM Q processor without errors mitigatation (b), and IBM q processor with errors mitigation (c) by application of  the map $\mathcal{E}_{z_{1}z_{2}}$ $i.e$ the eigenspace of $\sigma^{z_{1}} \otimes \sigma^{z}_{z_{2}}$ with $+1$ eigenvalue for $a_{z_{1}z_{2}} = 0$  }
  \label{fig:fig11}
\end{figure}

If the first and second qubits  share the same state, $i.e.$   $\ket{00q_{3}q_{4}}$ or $\ket{11q_{3}q_{4}}$, then, using $\mathcal{E}_{z_{1}}{z_{2}}$,  the initial mixed state is cooling into $+1$ eigenspace of the operator (stabilizer) $\sigma^{z}_{1} \otimes \sigma^{z}_{2}$ as shown in Fig. (\ref{fig:fig11} a)  using QASM simulator, and also the  noisy IBM Q quantum processor, without (\ref{fig:fig11}b)and with errors mitigation (\ref{fig:fig11} c) for the case when $a_{z_{1}z_{2}} = 0$.

\begin{figure}[ht]
  \includegraphics[width=1.0\linewidth]{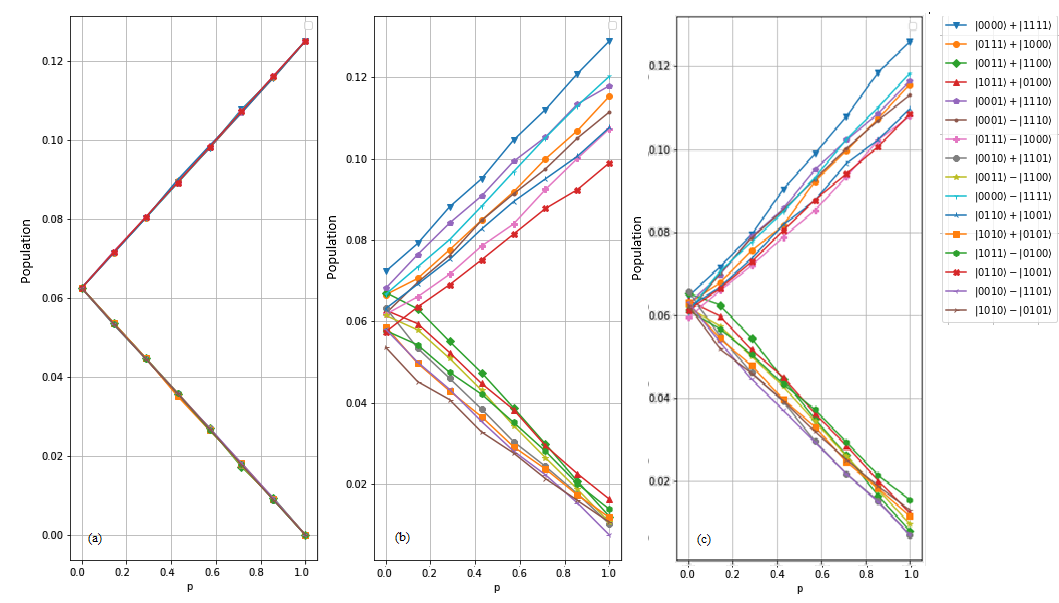}
  \caption{Simulation of the Markovian Lindblad master equation using QASM simulator (a), IBM Q processor without errors mitigatation (b), and IBM q processor with errors mitigation (c) by application of  the map $\mathcal{E}_{z_{2}z_{3}}$ $i.e$ the eigenspace of $\sigma^{z_{2}} \otimes \sigma^{z}_{z_{3}}$ with $+1$ eigenvalue for $a_{z_{2}z_{3}} = 0$}
  \label{fig:fig12}
\end{figure}

 However,  if the second and third qubits  share the same state, $i.e.$   $\ket{q_{1}00q_{4}}$ or $\ket{q_{1}11q_{4}}$, then, using $\mathcal{E}_{z_{2}}{z_{3}}$,  the initial mixed state is cooling into $+1$ eigenspace of the operator (stabilizer) $\sigma^{z}_{2} \otimes \sigma^{z}_{3}$ as shown in Fig. (\ref{fig:fig12}a)  using QASM simulator, and also the  noisy IBM Q quantum processor, without (\ref{fig:fig12}b)and with errors mitigation (\ref{fig:fig12}c) for the case when $a_{z_{2}z_{3}} = 0$

\begin{figure}[ht]
  \includegraphics[width=1.0\linewidth]{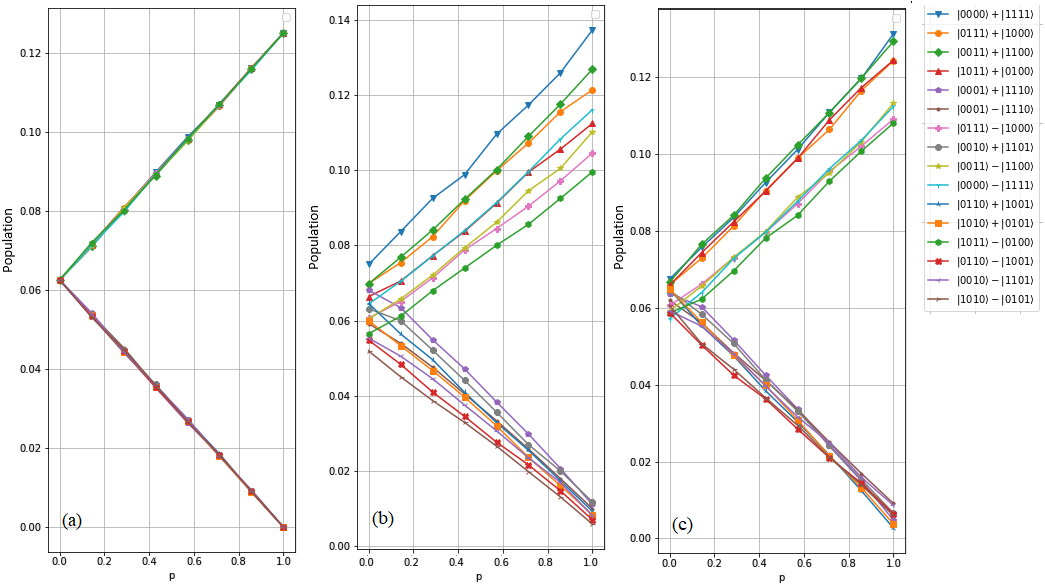}
  \caption{Simulation of the Markovian Lindblad master equation using QASM simulator (a), IBM Q processor without errors mitigatation (b), and IBM q processor with errors mitigation (c) by application of  the map $\mathcal{E}_{z_{3}z_{4}}$ $i.e$ the eigenspace of $\sigma^{z_{3}} \otimes \sigma^{z}_{z_{4}}$ with $+1$ eigenvalue for $a_{z_{3}z_{4}} = 0$}
  \label{fig:fig13}
\end{figure}

Likewise,  when  the third and fourth qubits  share the same state, $i.e.$   $\ket{q_{1}q_{2}00}$ or $\ket{q_{1}q_{2}11}$, then, using $\mathcal{E}_{z_{3}}{z_{4}}$,  the initial mixed state is cooling into $+1$ eigenspace of the operator (stabilizer) $\sigma^{z}_{3} \otimes \sigma^{z}_{4}$ as shown in Fig. (\ref{fig:fig13}a)  using QASM simulator, and also the  noisy IBM Q quantum processor, without (\ref{fig:fig13}b)and with errors mitigation (\ref{fig:fig13}c) for the case when $a_{z_{3}z_{4}} = 0$

In Fig. (\ref{fig:fig11}), Fig. (\ref{fig:fig12}), and Fig. (\ref{fig:fig13}), we have a noticeable affect of the   noises caused by the IBM Q processors  in the initial state mixed state’s preparation as well as the population of the targeted Bell’s state. 

\begin{figure}[ht]
  \includegraphics[width=0.5\linewidth]{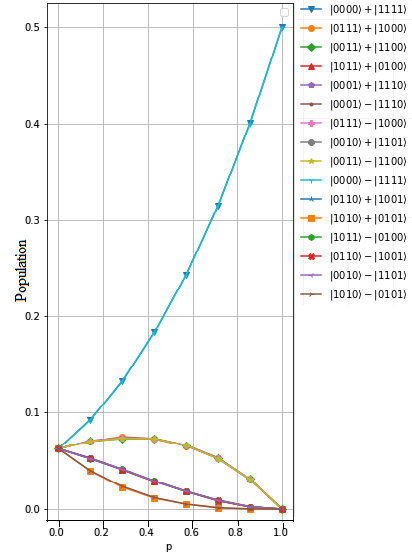}
  \caption{Simulation of the Markovian Lindblad master equation using QASM simulator using  the sequential  application of $\mathcal{E}_{z_{1}z_{2}}$, $\mathcal{E}_{z_{2}z_{3}}$, and $\mathcal{E}_{z_{3}z_{4}}$ maps in order to direct the initial state into   $1/\sqrt{2} \left(\ket{0000} \pm \ket{1111} \right)$ (when $a_{z_{1}z_{2}} = a_{z_{2}z_{3}}=a_{z_{3}z_{4}} = 0$ }
  \label{fig:fig14}
\end{figure}

Furthermore, Fig. (\ref{fig:fig14}) illustrates the population of the output state versus probability $p$ using  the sequential  application of $\mathcal{E}_{z_{1}z_{2}}$, $\mathcal{E}_{z_{2}z_{3}}$, and $\mathcal{E}_{z_{3}z_{4}}$ maps in order to direct the initial state into   $1/\sqrt{2} \left(\ket{0000} \pm \ket{1111} \right)$ (when $a_{z_{1}z_{2}} = a_{z_{2}z_{3}}=a_{z_{3}z_{4}} = 0$ for only  QASM simulator,  as the simulation using IBM Q processors in this case requires seven qubits where it is not available freely!

\onecolumngrid

\begin{figure}[ht]
  \includegraphics[width=0.5\linewidth]{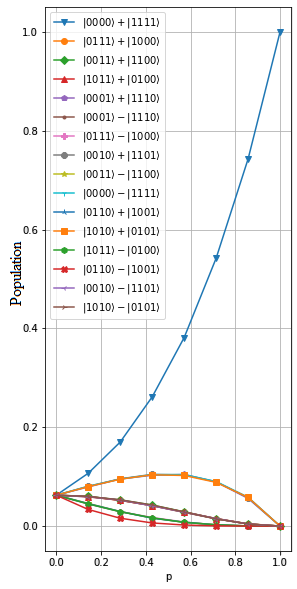}
  \caption{Simulation of the Markovian Lindblad master equation using QASM simulator using  the sequential  application of $\mathcal{E}_{z_{1}z_{2}}$, $\mathcal{E}_{z_{2}z_{3}}$, $\mathcal{E}_{z_{3}z_{4}}$, and  $\mathcal{E}_{x_{1}x_{2}x_{3}x_{4}}$ maps in order to direct the initial state into   GHZ state, $1/\sqrt{2} \left(\ket{0000} + \ket{1111} \right)$ (when $a_{z_{1}z_{2}} = a_{z_{2}z_{3}}=a_{z_{3}z_{4}} =  a_{x_{1}x_{2}x_{3}x_{4}}=0$}
  \label{fig:fig15}
\end{figure}

Finally, cooling into the GHZ state is completed    using the the map $\mathcal{E}_{x_{1}x_{2}x_{3}x_{4}}$ (when $a_{z_{1}z_{2}} = a_{z_{2}z_{3}}=a_{z_{3}z_{4}} =  a_{x_{1}x_{2}x_{3}x_{4}}=0$.

  \section{conclusion}
We demonstrate  preparing   of Maximally entangled state of two and four qubits dissipativley ,   by applying necessarily strong dissipative couplings using  the QASM simulator \cite{qiskit}, and free web based interface, IBM Quantum Experience (IBM QE) \cite{ibm},  to show how   noises  affect  the   state's preparation as well as the population of the targeted Bell's and GHZ states.   This  dynamics is frequently described  as continuous in time, described by many-body Lindblad master equations, cf. $e.g.$ \cite{gardiner}. The targeted  entangled  state is  required to be an eigenstate with respect to the quantum jump operators.  For future project research, we might increase the space  of the system and the environment, in this case the environment space contains controllable and uncontrollable state. 

\section*{Acknowledgement}
We acknowledge the use of IBM Quantum services for this work. The views expressed are those of the authors, and do not reflect the official policy or position of IBM or the IBM Quantum team.

\bibliography{bib.bib}
\bibliographystyle{apsrev4-1}

\end{document}